\documentclass[12pt]{article}

\usepackage{epsfig}

\def  \bcen   {\begin{center}}
\def  \ecen   {\end{center}}
\def  \beq    {\begin{equation}}
\def  \eeq    {\end{equation}}
\def  \beqa   {\begin{eqnarray}}
\def  \eeqa   {\end{eqnarray}}

\def  \unpart {{\cal U}}

\begin{document}

\renewcommand{\thefootnote}{\fnsymbol{footnote}}

\begin{flushright}
KEK-TH-1147, \\
arXiv:0705.1326 [hep-ph]
\end{flushright}
\vskip .5cm
\bcen
{\bf \Large Lepton Flavour Violation in Unparticle Physics} 
\vskip 1cm
{\large T. M. Aliev$^1$\footnote{taliev@metu.edu.tr}, 
A. S. Cornell$^2$\footnote{cornell@ipnl.in2p3.fr} and 
Naveen Gaur$^3$\footnote{naveen@post.kek.jp} }
\vskip .7cm
{\sl $^1$ Physics Department, Middle East Technical University, 06531 Ankara, Turkey, \\
$^2$ Universit\'e de Lyon 1, Institut de Physique Nucl\'eaire, Villeurbanne, France, \\ 
$^3$ Theory Division, KEK, 1-1 Oho, Tsukuba, Ibaraki 305-0801, Japan}
\ecen
\vskip .5cm

\begin{abstract}
Recently H. Georgi has introduced an unparticle $\unpart$ in order to
describe the low energy physics of a nontrivial scale invariant sector
of an effective theory. In this work we have explored the
phenomenology of an unparticle using the lepton flavour violating
decay $\mu^{-}\rightarrow e^{-} e^{+} e^{-}$, and found that the
branching ratio of this decay is strongly dependent on the scaling
dimension. 
\end{abstract}

\thispagestyle{empty}
\setcounter{page}{1}
%
In particle physics the concept of scale invariance is very predictive
in analysing the asymptotic behaviour of the correlation functions at
high energies. The  scale invariance at low energies is broken by the
masses of the particles. However, at a scale much beyond the Standard
Model (SM) the scale invariance is restored. Thus, if there is a scale
invariant sector, it can be experimentally probed at TeV
scale. Recently Georgi \cite{Georgi:2007ek,Georgi:2007si} observed
that such a conformal sector, due to the Banks-Zaks (BZ) fields
\cite{Banks:1981nn}, with scale dimensions $d_\unpart$, might appear
at the TeV scale. This sector at low energies he termed as
``unparticles'' ($\unpart$) \cite{Georgi:2007ek}, which are invisible
massless particles. In high energy collider experiments, like the LHC
or ILC, the missing energy distributions of various processes can
serve as a good test for the existence of these unparticles.   

\par The phenomenological implications of these unparticles in the
context of charged Higgs boson decays, anomalous magnetic moments, and
meson anti-meson mixings has been studied in reference
\cite{Luo:2007bq}. The possible effects of the $\unpart$ particle in
the missing energy signatures of $Z\rightarrow q\overline{q} \unpart$,
$e^{+} e^{-}\rightarrow\gamma \unpart$, and the monojet production at
hadronic collisions were performed in reference
\cite{Cheung:2007ue}. Note that there has also been another recent
study of unparticle physics within the context of B-decays
\cite{Chen:2007vv}, where the effects of $\unpart$ in hadronic Flavour
Changing Neutral Current (FCNC) processes was discussed. In another
work the constraints on unparticle physics from electron (g-2) to
positronium decays has been analysed \cite{Liao:2007bx}. 
In this work
we have tried to explore the phenomenology of an unparticle
($\unpart$) in the lepton flavour violating decay $\mu^{+}\rightarrow 
e^{-} e^{+} e^{-}$.  

\par Before commencing with the calculation a few words on this theory
are in order. At very high energy the theory contains the SM fields
and the fields of the scale invariant sector,  called BZ fields
\cite{Banks:1981nn}. The {\cal BZ} sector interacts with SM particles
through the exchange of a connector sector which has an high mass
scale $M_\unpart$. Below this mass scale the interaction between the
SM and the BZ sector manifests itself in terms of non-renormalizable
operators of the form; 
\begin{eqnarray}
O_{SM} O_{BZ}/M_{\unpart}^{k}  \ \ (k > 1) \,\,\, , \label{eq1}
\end{eqnarray}
where $O_{SM}$ and $O_{BZ}$ are the local operators constructed from
the SM and BZ fields. The renormalization effects in the scale
invariant sector induce the dimensional transmutation at the scale
$\Lambda_{\unpart}$ \cite{Coleman:1973jx}.  In the effective theory,
below the scale $\Lambda_\unpart$, of the BZ operators match onto
unparticle operators and non-renormalizable operators given by
equation (\ref{eq1}), where this matching gives rise to a new set of
interactions, having the form;  
\begin{eqnarray}
\frac{C_{\unpart}\Lambda_{\unpart}^{d_{BZ}-d_{\unpart}}}{M_{\unpart}^{k}}O_{SM} 
O_{\unpart} \,\,\, . \label{eq2} 
\end{eqnarray}
Note that $d_{BZ}$ and $d_\unpart$ are the scaling dimensions of the
$O_{BZ}$ and unparticle $O_\unpart$ operators respectively, and the
coefficient function $C_\unpart$ is fixed by the matching condition.  

\par Three unparticles operators with different Lorentz structures;
$O_\unpart$, $O_\unpart^\mu$ and $O_\unpart^{\mu\nu}$, are studied in
reference \cite{Georgi:2007ek}. The scale invariance was used to fix
the two point correlation functions of these unparticle
operators. Note that the abovementioned $O_\unpart^\mu$ and
$O_\unpart^{\mu\nu}$ were taken to be transverse. In reference
\cite{Georgi:2007si} the following set of effective operators, which
can have interesting phenomenological implications, were given; 
$$
\lambda_{0} \frac{1}{\Lambda^{d_{\unpart}}_{\unpart}} G_{\alpha\beta}
G^{\alpha\beta} O_\unpart \,, \quad
 \lambda_{1} \frac{1}{\Lambda^{d_{\unpart}-1}_{\unpart}} \overline{f}
\gamma_{\mu} f O^{\mu}_{\unpart}\,, \quad
\lambda_{2} \frac{1}{\Lambda^{d_{\unpart}}_{\unpart}} G_{\mu\alpha}
G^{\alpha}_{\nu} O_{\unpart}^{\mu\nu}  \,\,\, , etc. , 
$$
where $G^{\alpha\beta}$ is the gluon field strength tensor, $f$
denotes SM fermions and $\lambda_{i}$ are the dimensionless effective
coupling constants with $ \lambda_i = C_{O^{i}_{\unpart_{i}}}
\frac{\Lambda_{\unpart}^k}{M_\unpart^k}$. Note that where we have used
$i = 0, 1, 2$ we are referring to scalar, vector and tensor unparticle
operators respectively. The implications of these operators were
further analyzed in references
\cite{Georgi:2007si,Cheung:2007ue,Luo:2007bq}. In the analysis of
unparticle physics in $e^+ e^- \to \mu^+ \mu^-$, by Georgi
\cite{Georgi:2007si}, these operators were {\it assumed} to be {\it
  ``lepton-flavour-blind''}. In general these operators can be fermion
flavour dependent. In reference \cite{Chen:2007vv} the operators were
taken to be fermion flavour dependent and the analysis of these
operators in various B-decays was performed.  

\par For our study we will consider the following set of effective
interactions for the unparticle operators which have couplings to
leptons; 
\beq
\frac{c_V^{\ell \ell'}}{\Lambda_\unpart^{d_\unpart - 1}} \bar{\ell}
\gamma_\mu \ell'  O_\unpart  \ \ + \ \ 
\frac{c_A^{\ell \ell'}}{\Lambda_\unpart^{d_\unpart - 1}} \bar{\ell}
\gamma_\mu \gamma_5 \ell' O_\unpart \,\,\, .
\eeq
The scalar operator couples with ordinary SM fermions and in principle
can contribute to the problem under consideration. However, their
contributions are proportional to the fermion mass and are suppressed,
as we have electrons (positrons) in the final state. For this reason
we shall consider only the vector (axial-vector) unparticle operator
in this paper.     

\par The $\mu\rightarrow e^{-} e^{+} e^{-}$ decay is described by the
Feynman diagrams presented in figure (1). As such, in obtaining the
matrix element for the $\mu^{-} \rightarrow e^{-} e^{+} e^{-}$ decay
we need to know the propagator of the vector unparticles. This
propagator was first calculated in references \cite{Georgi:2007si,Cheung:2007ue}\footnote{Both the papers, references \cite{Georgi:2007si} and \cite{Cheung:2007ue}, have given
the explicit form of the propagator and appeared on the arXiv at the same time.} 
and has the explicit form; 
\begin{eqnarray}
D_{\mu\nu}(x)&=&\int d^4 x e^{i P x} <0 \mid
T[O^{\mu}_{\unpart}(x)O^{\nu}_{\unpart}(0)] \mid 0 > 
\nonumber \\
&=& \frac{i A_{d_{\unpart}}}{2 \sin d_{\unpart}\pi}
\left(-g_{\mu\nu}+\frac{P^\mu P^\nu}{P^2}\right)
(-P^2-i\epsilon)^{d_\unpart - 2} \,\,\, . \label{eq4} 
\end{eqnarray}
The vector operator is assumed to be transverse, that is $\partial_{\mu} O _{\unpart}^{\mu}=0$, where  
\begin{eqnarray}
A_{d_{\unpart}}=\frac{16 \pi^{5/2}}{(2 \pi)^{2 d_{\unpart}}}
\frac{\Gamma(d_{\unpart}+1/2)}{\Gamma(d_{\unpart}-1)\Gamma(2 d_{\unpart})}\,\,\, .
\label{eq9}
\end{eqnarray}
\begin{figure}[t]
\bcen
\epsfig{file=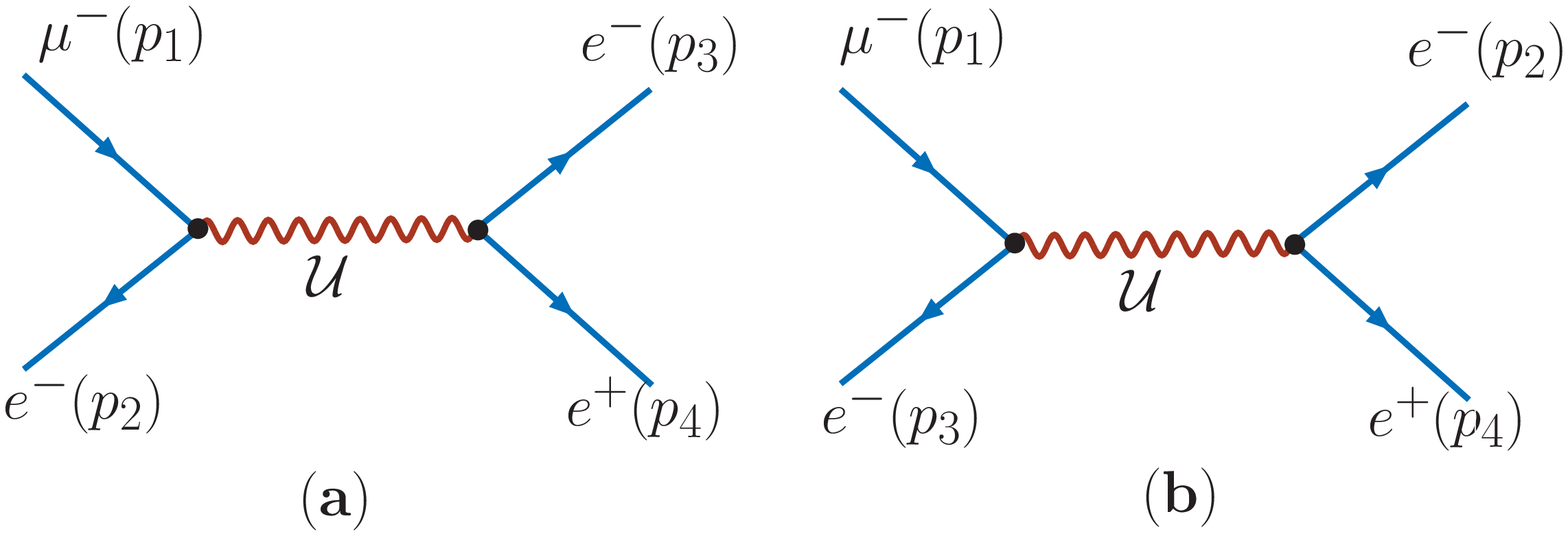,width=.8\textwidth}
\caption{\it Feynman Diagrams for $\mu^- \to e^- e^+ e^-$} 
\label{fig:1}
\ecen
\end{figure}
\noindent After making these deductions we can write the matrix element for the $\mu^{-} \rightarrow e^{-} e^{+} e^{-}$ decay as:  
\begin{eqnarray}
{\cal M} = {\cal M}_1 + {\cal M}_2 \,\,\, , \label{eq5}
\end{eqnarray}
with 
\begin{eqnarray}
{\cal M}_1&=& 
\bar{u}_2 \gamma_\mu \left(a_1 + a_2 \gamma_5\right) u_1 
\bar{u}_3 \gamma_\nu \left(a_3 + a_4 \gamma_5 \right) v_4
~~ \frac{A_{d_\unpart}}{\left(\Lambda^{d_\unpart - 1}\right)^2} 
\frac{\left(- g^{\mu \nu} + \frac{P^\mu P^\nu}{P^2}\right)}{2\ sin
  d_\unpart \pi} (- P^2 - i \epsilon )^{d_\unpart -2} \,\,\, ,
\label{eq6} \\
{\cal M}_2&=& -
\bar{u}_3 \gamma_\mu (a_1 +a_2\gamma_5) u_1 
\bar{u}_2 \gamma_\nu (a_3 +a_4 \gamma_5) v_4
~~ \frac{A_{d_\unpart}}{\left(\Lambda^{d_\unpart - 1}\right)^2} 
\frac{\left(-g_{\mu\nu}+\frac{Q_\mu Q_\nu}{Q^2} \right)}{2 \ sin
  d_\unpart \pi}  
\left( - Q^2 - i \epsilon \right)^{d_\unpart - 2}
\,\,\, , \label{eq7}
\end{eqnarray}
where $P=p_1-p_2=p_3+p_4$, $Q=p_1-p_3=p_2+p_4$ and $a_1 = c_V^{\mu e}, \ a_2 = c_A^{\mu e}, \ a_3 = c_V^{e e}, \ a_4 = c_A^{e e}$.

\par For a massless electron and positron the matrix elements ${\cal M}_1$ and ${\cal M}_2$ simplify greatly, reducing to the following forms: 
\begin{eqnarray}
{\cal M} _1&=& F_1 ~\overline{u}_2 \gamma_{\mu} (a_1 +a_2
\gamma_{5})u_1\overline{u}_3 \gamma^{\mu} (a_3 +a_4
\gamma_{5})v_4 \,\,\, ,
\nonumber \\
{\cal M} _2&=& - F_2 ~\overline{u}_3 \gamma_{\mu} (a_1 +a_2
\gamma_{5})u_1\overline{u}_2 \gamma^{\mu} (a_3 +a_4
\gamma_{5})v_4 \,\,\, , \label{eq10}
\end{eqnarray}
where 
\begin{eqnarray}
F_1&=&\frac{A_{d_{\unpart}}}{(\Lambda^{d_{\unpart}-1})^{2}
\sin(d_{\unpart}\pi)}(- P^2 - i \epsilon)^{d_{\unpart}-2} \,\,\, ,
\nonumber \\
\mathrm{and} \hspace{0.5cm} F_2&=&F_1 (P\rightarrow Q) \,\,\, . \label{eq8}
\end{eqnarray}
The matrix element squared, for the considered process, is then:
\begin{eqnarray}
|{\cal M}|^2=  |{\cal M}_1|^2+
|{\cal M}_2|^2
+ 2 Re ({\cal M}_1^{*} {\cal M}_2) \label{eq11}
\end{eqnarray}
with
\begin{eqnarray}
|{\cal M}_1|^2
&=& 32 | F_1 |^2
[(a_1^2+a_2^2)(a_3^2+a_4^2)\{(p_1.p_4)(p_2.p_3)+(p_1.p_3)(p_2.p_4)\}
\nonumber \\
&& \hspace{0.5cm}+4 a_1 a_2 a_3 a_4 \{(p_1.p_4)(p_2.p_3)-(p_1.p_3)(p_2.p_4)\}] \,\,\, ,
\nonumber \\
|{\cal M}_2|^2&=& 32 | F_2 |^2
[(a_1^2+a_2^2)(a_3^2+a_4^2)\{(p_1.p_4)(p_2.p_3)+(p_1.p_2)(p_3.p_4)\}
\nonumber \\
&& \hspace{0.5cm}+4 a_1 a_2 a_3 a_4 \{(p_1.p_4)(p_2.p_3)-(p_1.p_2)(p_3.p_4)\}] \,\,\, ,
\nonumber \\
Re ({\cal M}_1^{*} {\cal M}_2)
&=& 32 Re(F_1^* F_2)\{(a_1^2+a_2^2)(a_3^2+a_4^2)+4 a_1
a_2 a_3 a_4\}(p_1.p_4)(p_2.p_3) \,\,\, . \label{eq12}
\end{eqnarray}
The calculation will be performed in the centre of mass (CM) frame of
the outgoing electron and positron, as denoted by the momenta $p_3$
and $p_4$ respectively. In this frame we have;  
\begin{eqnarray}
p_1&=& (E_{\mu}\,,0\,,0\,,p\,) \,\,\, ,
\nonumber \\
p_2&=&(E_{e}\,,0\,,0\,,p\,) \,\,\, ,
\nonumber \\
p_3&=&(E_{e}^{\prime}\,,0\,,p^{\prime}\sin\theta\,,p^{\prime}\cos\theta\,) \,\,\, ,
\nonumber \\
p_4&=&(E_{e}^{\prime}\,,0\,,-p^{\prime}\sin\theta\,,-p^{\prime}\cos\theta\,) \,\,\, , \label{eq15}
\end{eqnarray}
with
\begin{eqnarray}
p&=& \frac{\sqrt{\lambda (m_{\mu}^2,m_{e}^2,s)}}{2\sqrt{s}}\simeq
\frac{m_{\mu}^2-s}{2\sqrt{s}}\,\, ,
\nonumber \\
p^\prime&=&\frac{1}{2}\sqrt{s-4 m_{e}^{2}}\simeq \frac{1}{2} \upsilon
\sqrt{s} \,\, ,
\nonumber \\
E_\mu&=&\frac{m_{\mu}^2-m_{e}^2+s}{2\sqrt{s}}\simeq
\frac{m_{\mu}^2 + s}{2\sqrt{s}}\,\, ,
\nonumber \\
E_e&=&\frac{m_{\mu}^2-m_{e}^2-s}{2\sqrt{s}}\simeq
\frac{m_{\mu}^2-s}{2\sqrt{s}}\,\, ,
\nonumber \\
E_e^\prime&=& \frac{\sqrt{s}}{2}\,\,\, , \label{eq16}
\end{eqnarray}
and where $s=(p_1-p_4)^2=(p_2+p_3)^2$, $v = \sqrt{1 - 4 m_e^2/s}$. In
this case $P^2=m_{\mu}^2+m_{e}^2-2 
E_{\mu}E_e+2p^2$ and $Q^2=m_{\mu}^2+m_{e}^2-2 E_{\mu}E_e^{\prime}-2p
p^{\prime}\cos\theta$.  

\par $|{\cal M }|^2$, after summing over all the final states and averaging over the initial states, is then used to calculate the decay widths using the general expressions: 
\begin{eqnarray}
d\Gamma &=& 
\frac{1}{2 E_\mu}  
\frac{d^{3}\overrightarrow{p}_2}{(2\pi)^3 2
E_2}\frac{d^{3}\overrightarrow{p}_3}{(2\pi)^3 2
E_3}\frac{d^{3}\overrightarrow{p}_4}{(2\pi)^3 2 E_4} (2 \pi)^4
\delta(p_1-p_2-p_3-p_4)  {1 \over 2} \ {1 \over 2} \ |{\cal M}|^2
\,\,\, , \label{eq13}  
\end{eqnarray}
\begin{equation}
\frac{d \Gamma}{d s d x} = \frac{1}{2^9 \pi^3} \frac{1}{\sqrt{s}} 
\left(\frac{m_\mu^2}{s} - 1 \right) \sqrt{1 - \frac{4 m_e^2}{s}} 
\ {1 \over 2} \ \ {1 \over 2} \ |{\cal M}|^2 \,\,\, ,
\label{eq14}
\end{equation}
where $x = cos(\theta)$ and $\theta$ is the scattering angle in the CM frame of the outgoing $e^+ e^-$. 

\begin{figure}[t]
\bcen
\epsfig{file=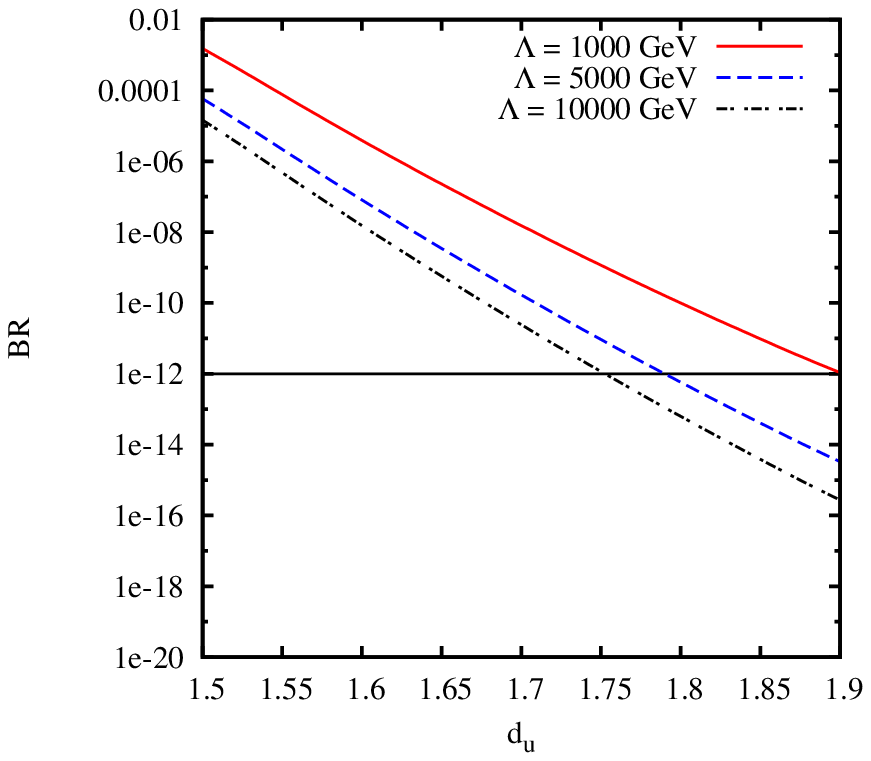,width=.6\textwidth}
\caption{\it Variation of the branching ratio with $d_\unpart$ for various values of $\Lambda_\unpart$. The other parameters are $a_1 = a_2 = 0.0001$ and $a_3 = a_4 = 0.01$}  
\label{fig:2}
\ecen
\end{figure}

\par From equation (\ref{eq14}) we calculate the branching ratio for the $\mu^{-}\rightarrow e^{-} e^{+} e^{-}$ decay, where we note that the present experimental limits are \cite{Yao:2006px}: 
$$
BR(\mu^- \to e^- e^+ e^-) < 1 \times 10^{-12} \,\,\, .
$$

\begin{figure}[t]
\bcen
\hskip -1.2cm \epsfig{file=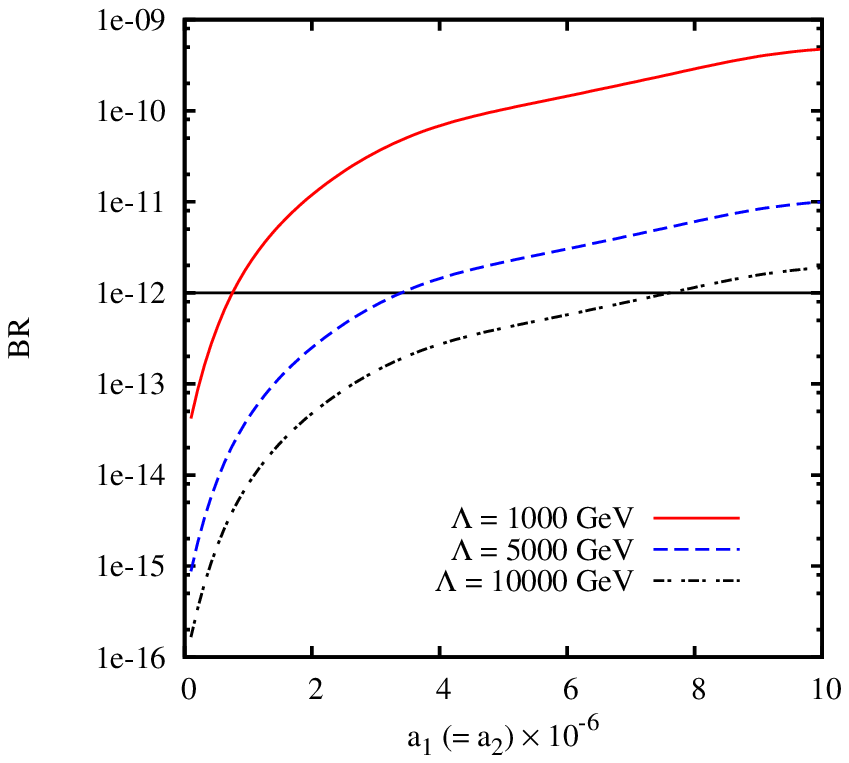,width=.6\textwidth} \hskip -2.5cm
\epsfig{file=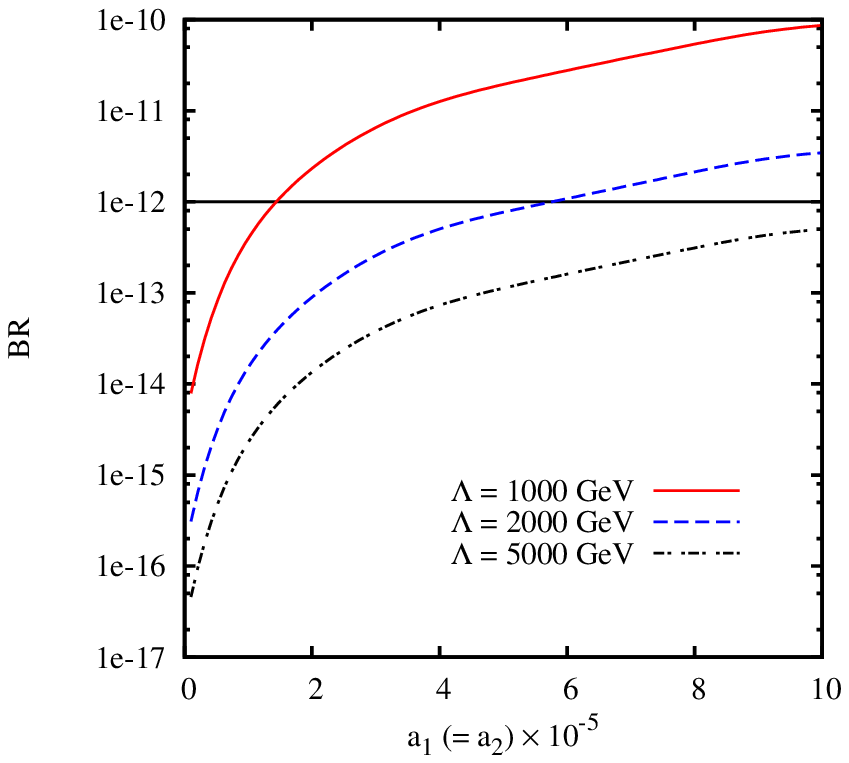,width=.6\textwidth}
\caption{\it Variation of the branching ratio with $a_1(= a_2)$ for
  various values of $\Lambda_\unpart$. The other parameters are 
(a) Left panel: $a_3 =  a_4 =   0.001$ and $d_\unpart = 1.6$,
(b) Right panel: $a_3 =  a_4 =   0.01$ and $d_\unpart = 1.8$.}
\label{fig:3}
\ecen
\end{figure}

\par In figure (\ref{fig:2}) we present the variation of the branching ratio as a function of $d_\unpart$ for various values of $\Lambda_\unpart$. As we can observe from this figure the branching ratio is very sensitive to the scaling dimension of the unparticle ($d_\unpart$). In figure (\ref{fig:3}) we have shown the dependence of the branching ratio of $\mu^- \to e^- e^+ e^-$ on the various $a_i$'s, where the coefficients $a_1$ and $a_2$ correspond to the lepton flavour violating (LFV) interactions. Again this has been done for different values of $\Lambda_\unpart$. 

\par As can be seen from these plots the LFV process $\mu^- \to e^- e^+ e^-$ is very sensitive to $d_\unpart$ and the LFV couplings $a_1$ and $a_2$. The same LFV couplings will be involved in other LFV processes, such as $\mu \to e \gamma$. It would, therefore, be interesting to explore the phenomenology of LFV unparticle operators on radiative LFV processes and their possible coorelation with $\mu^- \to e^- e^+ e^-$. It is very well known that SUSY predicts a very strong coorelation in the LFV processes $\mu \to e \gamma$ and $\mu^- \to e^- e^+ e^-$, where these coorelations tend to change substantially in the Little Higgs model with T-parity \cite{Blanke:2007db}. As such, it would be interesting to study the coorelations amongst various LFV modes in the presence of a scale invariant sector/unparticles. 

\par In conclusion, unparticle physics due to conformal invariance, might appear at the TeV scale and can lead to interesting phenomenological consequences which can be checked at low energy experiments. In this work we have studied the consequences of unparticle physics in the lepton flavour violating process  $\mu^{-}\rightarrow e^{-} e^{+} e^{-}$ and determined that the decay width is sensitive to the virtual effects of these unparticles.  

\section*{Acknowledgement}
We would like to thank Kingman Cheung and Yi Liao for their comments. 
The work of NG was supported by JSPS under grant no. P-06043.

\end{document}